\documentclass{osa-article}
\usepackage{amsmath}

\journal{osajournal}

\usepackage{booktabs}
\usepackage{tabularx}
\usepackage{array}
\newcolumntype{L}[1]{>{\raggedright\let\newline\\\arraybackslash\hspace{0pt}}m{#1}}
\newcolumntype{C}[1]{>{\centering\let\newline\\\arraybackslash\hspace{0pt}}m{#1}}
\newcolumntype{R}[1]{>{\raggedleft\let\newline\\\arraybackslash\hspace{0pt}}m{#1}}

\articletype{Research Article}

\begin{document}

\title{Potential of commercial SiN MPW platforms for developing mid/high-resolution integrated photonic spectrographs for astronomy}

\author{Pradip Gatkine,\authormark{1,*} Nemanja Jovanovic,\authormark{1}, Christopher Hopgood\authormark{2}, Simon Ellis\authormark{2}, Ronald Broeke\authormark{3}, Katarzyna Ławniczuk\authormark{3}, Jeffrey Jewell\authormark{4} and J. Kent Wallace\authormark{4}, Dimitri Mawet,\authormark{1,4}}


\address{\authormark{1}Department of Astronomy, California Institute of Technology, 1200 E. California Blvd., Pasadena, CA 91125, USA\\
\authormark{2}Australian Astronomical Optics, Macquarie University, 105 Delhi Rd, North Ryde, NSW 2113, Australia\\
\authormark{3}Bright Photonics BV, Horsten 1, 5612 AX Eindhoven, The Netherlands\\
\authormark{4}Jet Propulsion Laboratory, 4800 Oak Grove Drive, Pasadena, CA 91109, USA}

\email{\authormark{*}pgatkine@caltech.edu} 

\begin{abstract}
Integrated photonic spectrographs offer an avenue to extreme miniaturization of astronomical instruments, which would greatly benefit extremely large telescopes and future space missions. These devices first require optimization for astronomical applications, which includes design, fabrication and field testing. Given the high costs of photonic fabrication, Multi-Project Wafer (MPW) SiN offerings, where a user purchases a portion of a wafer, provide a convenient and affordable avenue to develop this technology. In this work we study the potential of two commonly used SiN waveguide geometries by MPW foundries, i.e.~square and rectangular profiles to determine how they affect the performance of mid-high resolution arrayed waveguide grating spectrometers around 1.5~$\upmu$m.  Specifically, we present results from detailed simulations on the mode sizes,
shapes, and polarization properties, and on the impact of phase errors on the throughput and cross talk as well as
some laboratory results of coupling and propagation losses.  From the MPW-run tolerances and our phase-error study, we estimate that an AWG with R $\sim$ 10,000 can be developed with the MPW runs and even greater resolving power is achievable with more reliable, dedicated fabrication runs. Depending on the fabrication and design optimizations, it is possible to achieve throughputs $\sim 60\%$ using the SiN platform. Thus, we show that SiN MPW offerings are highly promising and will play a key role in integrated photonic spectrograph developments for astronomy. 
\end{abstract}

\section{Introduction}

The ability to spectroscopically characterize objects is one of the most powerful diagnostic tools in an astronomer's arsenal. It can be used to determine the chemical composition and abundances of constituents, as well as the relative velocity of the object with respect to the observer, which can constrain its distance. By combining this data with imaging, it is possible to get a global understanding of a wide variety of astrophysical targets spanning from planets all the way up to the structure of the Universe. 

In the ultraviolet (UV) to mid-infrared spectral regions, astronomers typically rely on spectrographs constructed with classical optics such as  lenses, mirrors, gratings, and prisms. Prior to the invention of adaptive optics (AO), all ground-based telescopes operated in the seeing limit, providing a large  point spread function (PSF) to the slit of a spectrograph. The size of a seeing-limited spot is dependent on atmospheric turbulence, which varies temporally within a night and from site-to-site. The resolving power (R) of a seeing-limited spectrograph depends on the input spot size, which means the design and slit size of such a spectrograph needs to be optimized from site-to-site. In addition, the diameter of a seeing-limited  spectrograph collimator increases as a function of the telescope diameter, which has the knock on effect of increasing the size of subsequent optics and hence the volume, mass and cost of the spectrograph~\cite{BlandHawthorn2010-PIMMS}. The larger volume of a spectrograph adversely affects its thermal and mechanical stability, reducing the accuracy of the spectroscopic measurement. Therefore, instrument size and cost are critical issues for the next-generation extremely large telescopes (ELTs), which have primary mirror diameters of $>24$~m.

On the other hand, the size of a diffraction-limited instrument is independent of the site and telescope characteristics. This has two important properties: 1) a diffraction-limited spectrograph will have the same performance (R) at all telescopes and sites and 2) the instrument will be the most compact it can be for a given set of specifications~\cite{jovanovic2016-ESS}. These characteristics mean that diffraction-limited spectrographs could be replicated reducing the  development costs. In addition, very high R spectrographs (R > 100,000), designed for large telescopes (> 8 m), that would otherwise be extremely large ($>$ 15 m$^3$) and costly ($>$ \$10 million) \cite{quirrenbach2014carmenes}, can now be made compact and cost-effective. The advantages of diffraction-limited spectrographs have been noticed in the community.  RHEA~\cite{feger2016}, iLocator~\cite{crepp2016ilocater}, PARVI~\cite{gibson2019characterization}, and HISPEC/MODHIS \cite{Mawet2019High} have all adopted this architecture. Operating across $y$--$K$ bands, HISPEC/MODHIS will conduct $R>100,000$ science at both Keck and TMT Observatories (10 m and 30 m apertures respectively), while costing a fraction of first-light seeing-limited spectrographs under development for ELTs.

To enable a diffraction-limited instrument architecture however, the telescope itself must operate at the diffraction-limit, which is possible in three ways: 1) operating in space free from the effects of the atmosphere, 2) having a telescope diameter smaller than the Fried parameter, which implies a small diameter ($<0.5$ m) telescope, and/or 3) using an AO system to correct for the atmosphere. The third option is the most practical approach to combining a high-resolution spectrograph with a large collecting area telescope. Since AO systems are imperfect, and its beneficial to mount the spectograph off telescope, all solutions that rely on a diffraction-limited spectrograph design typically use single-mode fibers (SMF) to collect the light from the focal plane and route it to the instrument. The SMF acts as a spatial filter cleaning up any residual imperfections in the wavefront at the expense of coupling efficiency. As the coupling efficiency into a SMF is directly proportional to the Strehl ratio, the slowly improving Strehl ratio over the past 25 years has limited the applicability of this approach in the past. For low Strehl ratios, it is more efficient to capture the focused light into a multi-mode fiber (MMF) and use a  photonic lantern to adiabatically split the MMF into multiple diffraction-limited SMFs. However, this option requires using multiple diffraction-limited spectrographs for each SMF \cite{LeonSavalArgyrosBlandHawthorn, spaleniak2014multiband}. With the advent of ``extreme AO" systems~\cite{close2018SMR, beuzit2019sphere, macintosh2014GPI}, the Strehl ratio has finally reached $90\%+$ in the near IR (NIR) in better than median seeing conditions on bright stars, which means efficient direct  coupling to an SMF (without a photonic lantern) is finally a reality~\cite{jovanovic2017-EIL,bechter2020}. 

When operating at the diffraction-limit, it is possible to consider replacing the bulk optics with photonic circuits. In terms of spectrographs, there are numerous technologies that could be used including planar concave gratings, arrayed waveguide gratings (AWGs), and Fourier transform spectrometers, for example \cite{gatkine2019astrophotonic}. The additional benefit of using these devices is an even more compact instrument than can be achieved with bulk optics, reducing volume and mass, and increasing thermal and mechanical stability. At the same time, the single-mode regime offers immense flexibility to manipulate the light (before the spectroscopy step) such as adding in-line notch filters to specifically suppress atmospheric emission lines using waveguide/fiber Bragg gratings \cite{zhu2016arbitrary, ellis2020first}, precision wavelength calibrators \cite{obrzud2019microphotonic}, interferometers \cite{labadie2016astronomical, perraut2018single}, and photonic nullers to suppress unwanted starlight in a binary or a star-planet system~\cite{norris2020first}. Thus, the benefits of miniaturization and additional light-manipulation capabilities promote applications for both ground-based and space-borne instruments.  

Among the photonic spectrograph technologies, AWGs are the most promising for astronomy. AWGs have been extensively developed by the telecommunications industry and as such are the most mature. For this reason there have been previous efforts to characterize, optimize and build such devices into prototypical instruments for astronomy applications~\cite{cvetojevic2009,cvetojevic2012-DAW,cvetojevic2012-FSS,gatkine2017}, culminating in the demonstration of a highly efficient single-object spectrograph on an 8-m telescope~\cite{jovanovic2017-DEP}. These efforts have highlighted the potential for this technology in astronomy. The next steps are to customize the spectrograph characteristics to better match astronomical applications.  

There are many properties that can be optimized in such circuits including the R, the free-spectral range (FSR), the throughput, sensitivity to phase errors, spectral dropout, numerical aperture (NA) and mode-field diameter (MFD) for coupling into/out of devices, and polarization sensitivity. Although the science drives the requirements on these parameters, there are strong constraints imposed by the material platform and the manufacturing limitations. So it is critical to design the circuits with a holistic view of all of these terms in mind. 

To achieve the highest level of performance with the least amount of iterations, it is pragmatic to lean on commercial vendors with significant fabrication expertise with a given material platform. Commercial vendors understand that photonic prototyping can be very costly ($\gtrsim$ USD\$50k) and some offer Multi-Project Wafers (MPWs), which allow a user to purchase a section of a wafer, thus greatly reducing the cost of accessing the platform ($\sim$ USD\$10k to \$20k). This is ideal for developing circuit architectures and trading parameters of the design. These MPWs are offered on platforms like silicon nitride (SiN), Silicon-on-insulator (SOI), Indium phosphide (InP) and so on, but not silica-on-silicon (SOS), to the best of our knowledge. Of these platforms, SiN offers a moderate index contrast, which is ideal for coupling to/from \cite{zhu2016ultrabroadband}, while still being able to maintain small bend radii and hence compact circuits. There are two main classes of waveguide geometries offered by SiN MPW foundries, namely square and rectangular core shapes and both show great promise for enabling astronomers to rapidly prototype devices. 

In this body of work, we demonstrate through simulation, the potential of developing mid-to-high resolving power (R = 5k to 10k) AWGs based on SiN waveguides using the square and rectangular geometries of commercial MPWs. We explore the properties of the devices with realistic waveguide geometries and constraints imposed by the platforms in order to provide a potential future designer with a sense of the possibilities. In Section~\ref{sec:platforms}, we outline in detail some of the attributes of the two platforms. In Section~\ref{sec:simulations}, we describe the details of the design and simulation process. Section~\ref{sec:results} summarizes the main properties of the devices, while Section~\ref{sec:discussion} highlights the impacts of these properties. Section~\ref{sec:future} provides some ideas for future development of these platforms to enhance astronomical applications.

\section{Commercial platform attributes}\label{sec:platforms}

Exploiting commercial foundries for prototyping of photonic circuits brings significant experience and process control, but can be a costly approach. As outlined in the previous section, this is where MPWs can help. SiN is the optimal platform to consider for astronomical applications thanks to its favorable throughput compared to others (e.g.~SOI, InP, etc.). In addition, SiN also has the following beneficial properties: a large spectral transparency window allowing a broad operational band ($\lambda$ = 400$-$2400 nm) \cite{blumenthal2018silicon}, a higher index contrast compared to silica-on-silicon allowing tight bends with $<$ 100~$\upmu$m radius with low radiation losses \cite{bauters2011ultra}, low propagation losses, adjustable MFDs to efficiently couple to optical fibers, and a relatively high $dn/dT$ for realizing compact thermal phase shifters \cite{ arbabi2013measurements, elshaari2016thermo}. The high index contrast/small bend radius makes it possible to design very compact AWGs, which in turn minimizes propagation losses as well. Having the flexibility to adjust the MFD of the guides at the interfaces to the chip is also critical to minimizing coupling losses and maximizing the overall system throughput. 

There are numerous commercial foundries that can support MPW runs in SiN including but not limited to Ligentec~\footnote{https://www.ligentec.com/}, LioniX~\footnote{https://www.lionix-international.com/}, LETI~\footnote{https://europractice-ic.com/}, and AMF. One differentiating feature of the offerings is the waveguide geometry. For example, Ligentec specializes in making square waveguides with dimensions of $800\times800$~nm, while most other vendors focus on rectangular waveguides with single and dual stripes with typical thicknesses of 50$-$250 nm/stripe. Indeed, making thick layers of SiN is very challenging as the structures typically suffer large stresses and crack, but Ligentec have developed a proprietary approach that circumvents this allowing for square waveguides with ultra-low propagation losses ($\sim$0.1~dB/cm). Square waveguides offer polarization insensitivity and hence, are better-suited for the unpolarized sources in astronomy. We will explore both options in the simulations that follow.

\section{AWG Design and Simulations}\label{sec:simulations}
In this section, we introduce the AWG properties and describe our design and simulation schemes for the AWG spectrographs. 

\subsection{AWG properties}\label{subsec:AWG_prop}

\begin{figure}[h!]
\centering\includegraphics[width=0.99\textwidth]{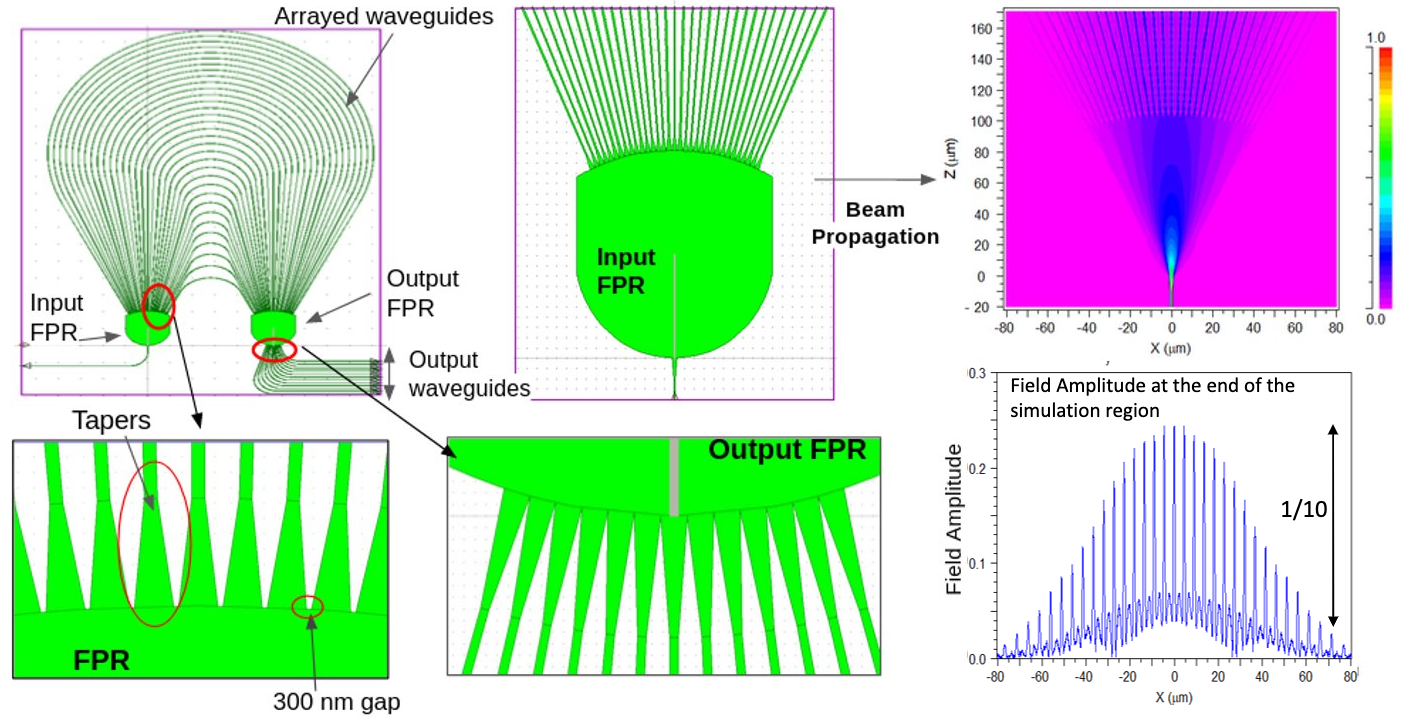}
\caption{\footnotesize The structural components of an AWG. \textbf{Top left:} AWG schematic showing the full CAD. 
\textbf{Top center:} The input FPR simulated with 3D EIM. There are tapers at the FPR-waveguide interface on both sides. The length of waveguide array simulated with EIM was up to the point where the lateral separation of the waveguides was $>$ 2$\times$MFD. \textbf{Top right:} The electric field amplitude obtained through the 3D EIM beam propagation simulation of the input FPR. \textbf{Bottom right:} The electric field amplitude at the end of the simulated region showing the Gaussian envelope of the illumination pattern of the input waveguide (plotted down to 1/10$\mathrm{^{th}}$ of the maximum field amplitude) as sampled by the waveguide array (seen as narrower peaks). \textbf{Bottom left:} The interface between the FPR and the array of waveguides showing the linear tapers and the 300~nm gap that has been introduced in to simulate the MPW fabrication tolerance. \textbf{Bottom center:}  The interface between the output FPR and the output waveguides (receiver channels) showing the linear tapers touching each other in order to minimize the spectral dropout.}\label{fig:AWG_schematic}
\end{figure}

The general scheme and terminology of an AWG-based spectrograph are explained in \cite{gatkine2016development} and Fig. \ref{fig:AWG_schematic}. To reiterate, an input waveguide illuminates an array of waveguides through a region called the input free propagation region (FPR). The ends of the FPR follow a Rowland curvature. The output of this FPR has an array of single-mode waveguides that collect the diffracting beam. The array of waveguides is designed to introduce a phase slope, with a constant path length difference between adjacent waveguides. The resultant phased array creates an interference pattern with constructive interference (i.e.~peak) for different wavelengths at different spatial points along the Rowland curvature of the output FPR (i.e.~the focal plane). This is where the dispersed spectrum is formed that is then sampled by discrete output waveguides or by directly imaging the focal plane on a detector~\cite{cvetojevic2012-DAW}. In the case of discrete sampling, the spectral separation between the neighboring output waveguides ($\delta\lambda$) is called the channel spacing.

The key properties of an AWG transmission response include: FSR, insertion loss, crosstalk, and spectral dropout. They are briefly described below. For further details about each of these properties, we direct the reader to \cite{okamoto2010fundamentals}.

\begin{figure}[h!]
\centering\includegraphics[width=0.99\textwidth]{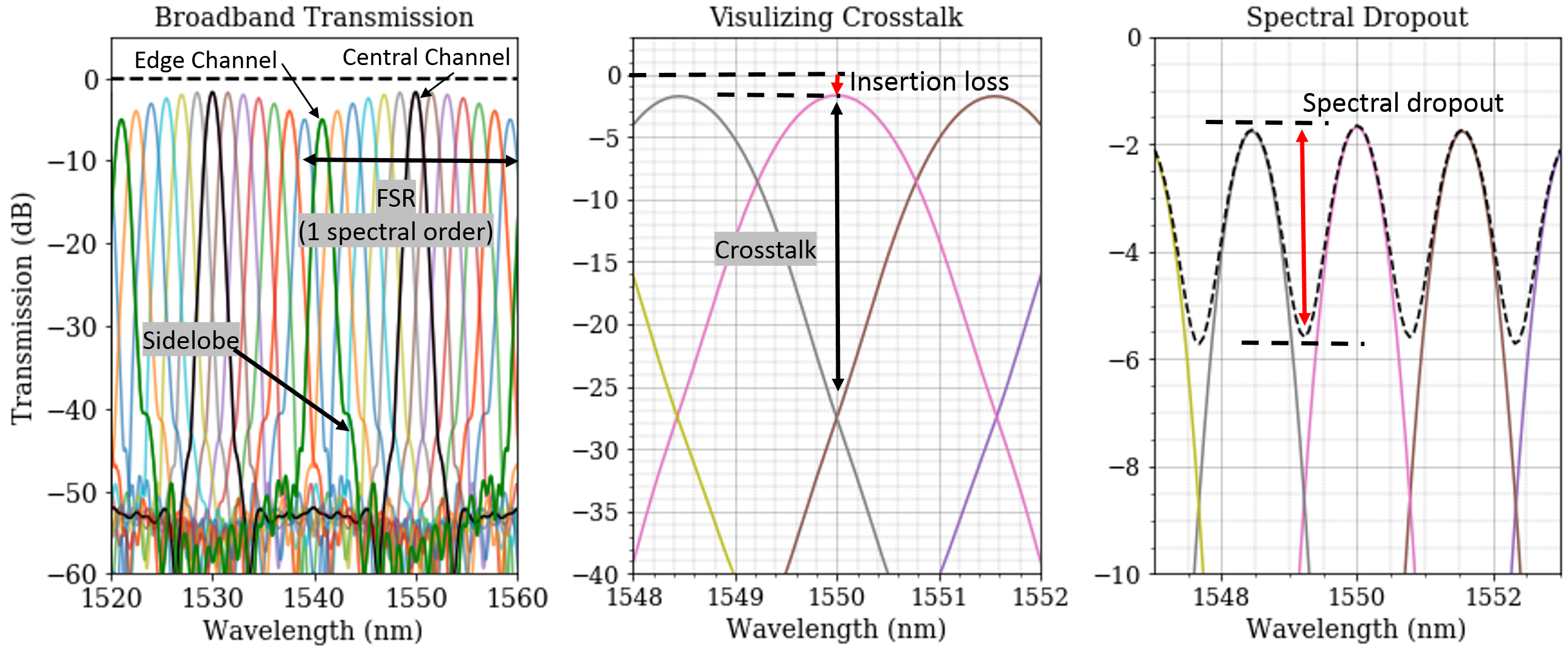}
\caption{\footnotesize A representative AWG spectral response showing the key spectral features of a R=1000, FSR=20~nm device with a center wavelength of 1550~nm. 
\textbf{Left:} The transmission response of the AWG showing two spectral orders (1520-1560 nm) with spectral orders spanning 20~nm (i.e.~the FSR). The corresponding central and edge channels in the two spectral orders are highlighted in black and green, respectively. 
\textbf{Center:} The crosstalk and insertion loss for the central channel are shown here.  
\textbf{Right:} The dashed black line is the summed spectrum over all output channels. The spectral dropout around the central channel is indicated by the red arrow. 
}\label{fig:AWG_spec_param}
\end{figure}

\textbf{Free spectral range:} Figure~\ref{fig:AWG_spec_param} shows the transmission response of an AWG with a R=1000 (i.e.~channel spacing = 1.55~nm) and a FSR = 20~nm. Each peak within a FSR indicates one output waveguide. Note that multiple spectral orders (i.e.~FSRs) across the range of operation of the AWG (say, 1450$-$1650~nm) spatially overlap at the output FPR (and thereby, at the output waveguides). For instance, the central output waveguide will have the spectral response shown in black (i.e.~peaks for $\lambda$ = 1530 nm, 1550 nm, 1570 nm and so on) due to the spatially overlapping spectral orders.

\textbf{Insertion loss:} The insertion loss includes the losses at the waveguide-FPR interfaces (due to mode-field mismatch at the FPR-taper interfaces, gaps between the tapers, and the truncation of the illumination pattern of the input waveguide) and the light lost to higher spatial orders in an AWG (the spatial orders beyond the region sampled by the output waveguides). The insertion loss is shown in Fig.~\ref{fig:AWG_spec_param}b. Note that the insertion losses of the edge channels for a given order are higher than those of the central channels as can be seen in Fig.~\ref{fig:AWG_spec_param}a. The envelope of AWG peaks is determined by the far field pattern of the waveguides illuminating the output FPR. 

\textbf{Crosstalk:} We define crosstalk as the difference between the peak transmission for a given channel and the transmission of its neighboring channel at that peak wavelength. It is a measure of the contrast ratio or sharpness of the AWG peaks. The electric field at an output channel is the image of the electric field of the input waveguide modulated by the sampling by the array of waveguides. The sampling function introduces sidelobes in the spectral profile, which contribute to a higher spectral leakage. In addition, broader tapers at the output FPR (e.g.~touching tapers shown in Fig.~\ref{fig:AWG_schematic}, bottom-center panel) broaden the spectral profile of an output channel, thus leading to a higher crosstalk. In astronomical spectroscopy, this is analogous to the line spread function \cite{casini2014instrument}. For astronomical sources, it is more critical to collect all the light at the output FPR than the spectral purity of the channels as long as the spectral leakage is $<$ 30-40\%  (or crosstalk $>$ 4$-$5 dB) and the line spread function is well-characterized. Therefore, touching tapers at the output FPR are better suited for astronomy.

\textbf{Spectral dropout:} Another requirement in astronomical spectroscopy is the ability to reconstruct a continuous spectrum to ensure a complete picture of the target and that lines of interest are not missed. However, in a typical AWG, the dispersed spectrum at the output FPR is sampled by discrete output waveguides (or receiving channels), thus giving rise to a discreteness in the spectral response. This can be seen in Fig.~\ref{fig:AWG_spec_param}b where each channel (indicated with a different color) has a peak at one wavelength and the response tapers off as we go away from that peak wavelength. If we sum all channels together, we get an idea of the total throughput of the device. This gives rise to a wavy response as shown in Fig.~\ref{fig:AWG_spec_param}c. Thus, there is an increased loss in the spectral regions between the peak wavelengths of any two spectral channels. We define this loss as the spectral drop-out, which occurs due to the discretization of the output FPR (or the focal plane). There is no such dropout if we directly image the focal on a detector. Here we have only defined the magnitude of the drop-out, but the width is equally as important and depends on the device design. We only focus on the amplitude in this work.

\subsection{Waveguide Geometry}
\begin{figure}[h!]
\centering\includegraphics[width=0.99\textwidth]{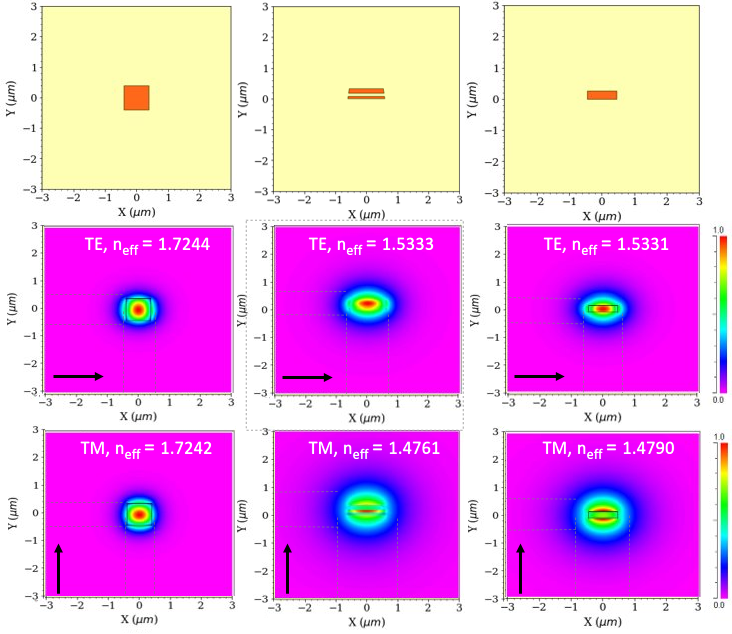}
\caption{\footnotesize Waveguide cross-sectional geometries and mode profiles of the a) square waveguide (Ligentec-like), b) double-stripe waveguide (Lionix MPW profile), c) rectangular waveguide (equivalent to Lionix MPW profile from the perspective of effective index and the mode profiles of both polarizations). The black arrows indicate orientation of electric field (E$_{\rm x}$ and E$_{\rm y}$) in the profiles shown here. Dashed lines indicate the 1/e$^{2}$ intensity locations and hence the size of the MFD.}\label{fig:mode_profiles}
\end{figure}

Given the range of possibilities, we will focus our simulation efforts here on the two main kinds of the commercial MPW geometries available, namely a square (Ligentec-like) and rectangular (LioniX-like) geometry and simulate a range of AWG devices using these to understand how this impacts the various properties of the spectrographs. Figure~\ref{fig:mode_profiles} shows the cross-sectional geometries of the waveguides utilized in all the simulations. The left panel describes the square waveguide and is representative of the Ligentc MPW offering. In reality the side walls of the guides are inclined at $89^{\circ}$. However, we simulate them as perfectly vertical (i.e.~$90^{\circ}$) for ease of simulation without altering the mode or the effective index in any significant way. The right panels show the rectangular (Lionix-like) waveguide geometry. LioniX's MPW offering uses a dual stripe approach with a top and bottom stripe thickness of 175 and 75 nm respectively, and a separation between the two stripes of 100 nm (as shown in the center panels of Fig~\ref{fig:mode_profiles}; for a detailed summary, see the Triplex platform in \cite{roeloffzen2018low}). However, to eliminate the need for full 3D simulation, we identified an equivalent rectangular waveguide geometry providing a similar mode size and effective index through an iterative process. The rectangular analog is 900 nm wide and 263 nm thick as shown in Fig.~\ref{fig:mode_profiles} - right panels. This single stripe rectangular analog was used in all further simulations. Note that the advantage of a double stripe geometry is that it offers a relatively high confinement factor without requiring thick SiN deposition for either stripes, thus avoiding the material stresses (and stress-induced microcracks \cite{irene1976residual}).


\subsection{AWG Design: Spectral parameters}

In order to explore AWG designs employing the two geometries on equal footing, we fixed the  channel spacing (i.e.~a discrete resolution element) of the AWG and the FSR. Astronomical spectrographs come in a range of resolving powers depending on the science case. For instance, a lower resolving power (R $\sim$ 1000) is useful for taking the spectra of faint objects or for measuring galaxy redshifts, while high resolving power (R $>$ 10,000) is useful for measuring the composition and kinematics of galactic flows or composition of exoplanet atmospheres~\cite{wang2017-OEH}. We therefore choose three resolutions (R $\sim$ 1000, 5000, 10000) to explore the parameter space of interest. Even higher R is of interest in astronomy, but we start our efforts at more modest values for simplicity. Here, we define the resolving power as $\lambda/\delta\lambda$ where $\delta\lambda$ is the discrete channel spacing of the AWG (discretized due to output waveguides).  

We choose the near-IR band around 1550 nm for simulating these AWGs for several reasons. First, the SiN platform and foundry processes are optimized around 1550 nm since this is the waveband of interest for telecommunication applications. Second, for galaxies and quasars in the early universe (cosmological redshift $>$ 6), the rest-frame optical and ultra-violet wavebands which carry the most useful spectral lines tracing the multi-phase gas (e.g.~C$^{3+}$, Mg$^{+}$, Fe$^{+}$) are all redshifted to the near-IR \cite{shen2019gemini}. Third, terrestrial exoplanet atmospheres should contain molecular species such as H$_2$O, CH$_4$, CO$_2$, which have spectral features in the near-IR \cite{wang2017-OEH}. Therefore we choose the center wavelength of the AWG as 1550 nm.    

A FSR of 20 nm is chosen to limit the number of waveguides in the array, and therefore, keep the footprint of the AWG within a standard reticle area (typically 100$-$200 mm$^2$). At the same time, this FSR is large enough to cover the whole astronomical H-band (1490$-$1800 nm) in 15 spectral orders, thus requiring only a low-resolution cross-dispersion \cite{cvetojevic2012-FSS, gatkine2020development}.

\subsection{AWG Design: Structural parameters}
Given the spectral parameters and the waveguide geometry, we design the AWGs for each case using the design algorithm described in \cite{smit1996phasar}. As explained earlier, we fix the center wavelength ($\lambda_{0}$) at 1550 nm and the FSR at 20 nm. The spectral order ($m$) is determined in each case using the equation: $FSR$ $\approx$ $\lambda_{0}/m$ $\times$ $n_{\rm eff}/n_{\rm group}$, where $n_{\rm eff}$ and $n_{\rm group}$ are the effective and group indices of the waveguide, respectively. The path difference between the adjacent waveguides in the array is determined as: ($\Delta L$ = $m\lambda_{0}/n_{\rm eff}$). The number of output channels (or discrete resolution elements) is determined by the FSR and channel spacing ($\delta\lambda$, or the spectral width of one resolution element) as $N_{\rm chan}$ = $FSR/\delta\lambda$. 

The AWG is designed to keep the nominal crosstalk level (see definition above) to less than $-$20 dB for the central channel, which dictates both the number of waveguides in the array and the length of the FPR. Given the different waveguide geometries we are considering here, it is important to ensure that we maintain the same number of waveguides in the array for a similar resultant resolving power and a fair comparison. This was achieved by adjusting the length of the FPR for each geometry until the outermost waveguide in the array captured no more than $\mathrm{\sim 1/10^{th}}$ of the flux of the central waveguide (refer to the right panels of Fig.~\ref{fig:AWG_schematic}). This criterion also ensured that most of the light was captured by the array. The key structural properties of the AWGs are summarized in Table \ref{tab:AWG_prop_table}.

We further added adiabatic linear tapers at all the four slab-waveguide interfaces (two on the input and two on the output FPRs) with a width of 3$\times$W and a length of 10$\times$W (where W is the waveguide width) to ensure a smooth transition of the slab mode to the waveguide mode (and vice versa). A spacing of 300 nm was added between the tapers at the interface between input/output FPRs and the waveguide array to account for the smallest feature size limitation of the lithography process used in the commercial MPW runs. We do not enforce this limitation at the interface of the output FPR and the output waveguides (see Fig. \ref{fig:AWG_schematic} bottom-center panel) since the output waveguides could be polished off to directly image the continuous spectrum on a detector. The touching tapers along the Rowland curvature of the output FPR ensure that most of the spectrum is captured without significant dropout.

The minimum allowed radius of curvature for the waveguides in the array depends on the specific waveguide geometry to maintain low bend losses. Foundries have experimentally determined that if the radii of the waveguides are kept above this minimum criterion, the bend losses can be $<$0.5 dB/cm. To operate in this low-loss regime, a minimum bend radius of 20~$\upmu$m if only the TE polarization is considered or 150~$\upmu$m if both polarizations are required, is needed for our square waveguides (Ligentec-like; 0.8 $\times$ 0.8 $\upmu$m). Similarly, a minimum bend radius of 80~$\upmu$m is needed for the TE polarization for our rectangular waveguides (Lionix-like, 0.9 $\times$ 0.26~$\upmu$m; only TE polarization is considered since the rectangular waveguides are birefringent).    

\begin{table}[ht]
\footnotesize
\begin{center} 
\begin{tabular}{|C{3.2cm}|C{2cm}|C{2cm}|C{2cm}|} 
\hline
\rule[-1ex]{0pt}{3.5ex}  \textbf{Geometry = 0.8 $\times$ 0.8 $\upmu$m} & \textbf{No. of waveguides} & \textbf{Length of FPR ($\upmu$m)} & \textbf{$\Delta$L ($\upmu$m)} \\
\hline
\rule[-1ex]{0pt}{3.5ex}  Res 1000 & 33 & 99.9 & 57.3 \\
\rule[-1ex]{0pt}{3.5ex}  Res 5000 & 163 & 506.6 & 58.0 \\
\rule[-1ex]{0pt}{3.5ex}  Res 10000 & 323 & 1013.3 & 58.0 \\
\hline
\rule[-1ex]{0pt}{3.5ex}  \textbf{Geometry = 0.9 $\times$ 0.26 $\upmu$m} & & & \\
\hline
\rule[-1ex]{0pt}{3.5ex}  Res 1000 & 33 & 110.7 & 65.9 \\
\rule[-1ex]{0pt}{3.5ex}  Res 5000 & 163 & 553.5 & 65.9 \\
\rule[-1ex]{0pt}{3.5ex}  Res 10000 & 323 & 1104.3 & 64.7 \\
\hline
\end{tabular}
\caption{\label{tab:AWG_prop_table} \footnotesize Structural properties of the AWGs discussed in this paper for both the square and rectangular geometry. The spectral order ($m$) is 64 in both the designs. }
\end{center}
\end{table}

\subsection{Simulations}
We simulate the designs obtained in the step above using RSoft CAD\cite{rsoft200layout} and the AWG utility of the BeamPROP module\cite{BeamPROP}. We use the 3D effective index method (EIM) to simulate the propagation from the input waveguide to the FPR and into the waveguides in the array until the modes in the array of waveguides are well-separated (by at least 2 $\times$ the MFD) to ensure proper decoupling of the electric field. After this simulation, the phase increment and phase errors are added to the waveguides in the array. Finally, the beam propagation in the output FPR is simulated using the 3D EIM to construct the spectrum received at the output waveguides. The AWGs are designed for the TE mode and are simulated by illuminating both TE and TM modes to understand the polarization-dependent response of the designs.   

\subsection{Simulation of phase errors}
The array of waveguides needs to be phased precisely before illuminating the second FPR in order to produce a spectrum. Any phase errors can deteriorate the sharpness of the peaks and will therefore, determine the contrast between the spectral peaks and the side lobes. This in turn determines how high a resolution spectrograph can be built using a fabrication platform, as discussed below. 

Phase errors in the array could arise from side wall imperfections, defects, thickness variations in the deposition processes, stitching errors in case of e-beam lithography or if using multiple reticles for a single device in case of stepper lithography, and so on \cite{stoll2020performance}. To simulate the impact of these imperfections, we added phase errors to our simulations. This is achieved by altering the phase in each waveguide of the array by a random number uniformly distributed between 0 and a maximum phase error value ($\Delta \phi$ in degrees). The resulting simulated spectral response for each AWG (with different spectral parameters and waveguide geometries) is discussed in section \ref{sec:results} through various metrics such as throughput and crosstalk as a function of maximum phase error. These results are helpful in determining the limits on the spectral parameters that can be achieved given a phase error tolerance for the fabrication process and the desired performance. In section \ref{subsec:phase_err_fabrication}, we estimate how these phase-error tolerances translate into geometrical tolerances of width and height fluctuations for different AWG devices and thereby discuss the potential limitations imposed by the fabrication processes on building high-resolution devices.

\section{Results}\label{sec:results}

\subsection{Spectral profile dependence on phase errors}
\begin{figure}[h!]
\centering\includegraphics[width=0.99\textwidth]{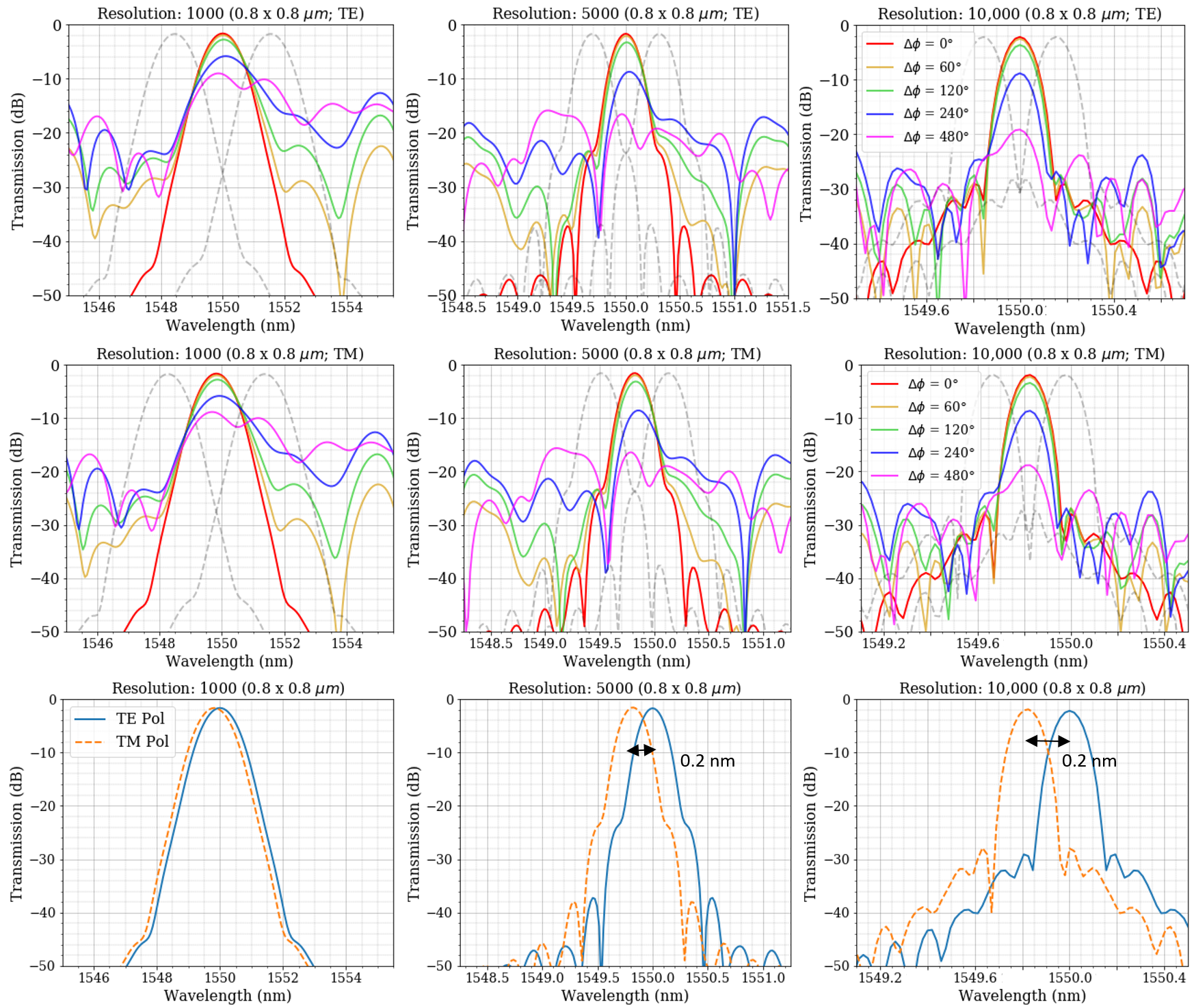}
\caption{\footnotesize Spectral response of a single channel of AWGs designed using square waveguides (Ligentec-like), as a function of resolving power, for TE and TM polarization's. The impact of phase errors on the spectral profile is shown in all cases. The bottom panel shows the wavelength offset between the TE and TM polarizations. }\label{fig:LGT_spectra}
\end{figure}

\begin{figure}[h!]
\centering\includegraphics[width=0.99\textwidth]{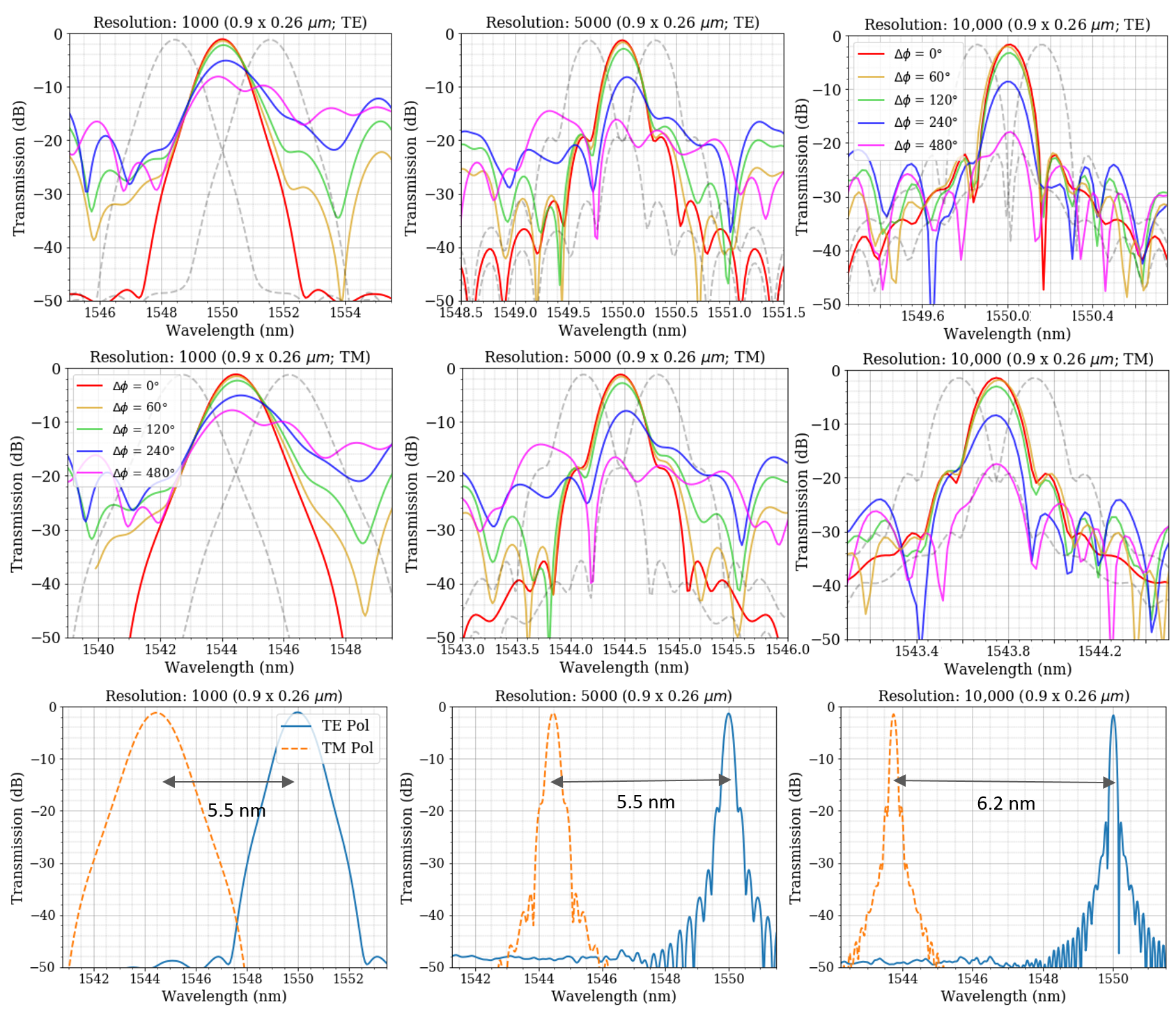}
\caption{\footnotesize Spectral response of a single central channel of AWGs designed using rectangular waveguides (LioniX-like), as a function of resolving power, for TE and TM polarization's. The impact of phase errors on the spectral profile is shown in all cases. The bottom panel shows the wavelength offset between the TE and TM polarizations.}\label{fig:LiX_spectra}
\end{figure}

We start our investigation by examining the impact of phase errors on AWG properties. Figures~\ref{fig:LGT_spectra} and \ref{fig:LiX_spectra} show the profile of a single spectral channel (the central channel in an FSR), as a function of resolving power and phase error, for TE (in the top row) and TM polarizations (in the middle row)  for the square and rectangular waveguides respectively. The dashed profiles show the neighboring output channels (only for zero phase error) to provide an estimate of the spectral degradation and crosstalk with phase error. The first thing to notice is that in the cases where there are no phase errors, a discrete spectral channel is formed, indicating that both waveguide geometries can support AWGs with a range of resolving powers assuming perfect fabrication. 

It can be seen that the profiles of the spectral channels for TE and TM polarizations are near identical in the case of the square waveguides (Fig. \ref{fig:LGT_spectra}). In addition, there is only a slight spectral offset of 0.2 nm between the wavelength of the central channels for TE and TM polarizations. This is unsurprising given the symmetry of the waveguide geometry. 
In comparison, for the rectangular waveguides, the sidelobes of TM polarization are broader than those for TE (in Fig. \ref{fig:LiX_spectra}, compare the side lobes of the red channels in each plot), which also results in a slightly greater crosstalk with the adjacent channels for the TM mode. Unlike the square waveguides, the rectangular ones have a large central wavelength offset between the two polarizations ($\sim$ 5.7 nm $+$ 2$\times$FSR). This is expected, given the significantly broader and lower effective-index mode profile for the TM mode as opposed to the TE mode in the rectangular waveguide (as shown in Fig.~\ref{fig:mode_profiles}). In other words, the TE and TM modes at a single wavelength will land in different regions of the output FPR and couple into different output waveguides. Therefore, the rectangular waveguides are well-suited in the cases where only a single polarization is fed to the AWG, or the polarizations are separated and then fed to two different AWGs (with rectangular waveguides) optimized for the respective polarizations. 


Finally, as the phase error present in the array of waveguides is increased, the peak transmission of the spectral channel reduces (i.e.~the insertion loss increases), and the side lobes get stronger. This is because as the phase error increases, the sharpness (or the coherence) of the constructive interference peaks diminishes. The reduced efficiency is a result of the light of a given wavelength being spread over a larger region at the output of the second FPR and hence coupling into a larger number of output waveguides, increasing the strength of the side lobes at the same time. It can be seen that for phase errors up to 120$^{\circ}$ (green curves), despite the elevated side lobes, the primary spectral channel profile is retained with little reduction in peak efficiency. In section \ref{subsec:phase_err_fabrication}, we estimate the corresponding width or height errors permitted for devices designed with each waveguide geometry to result in the level of phase error shown in the plots above.  

\subsection{Insertion loss and crosstalk dependence on phase errors}
\begin{figure}[h!]
\centering\includegraphics[width=0.99\textwidth]{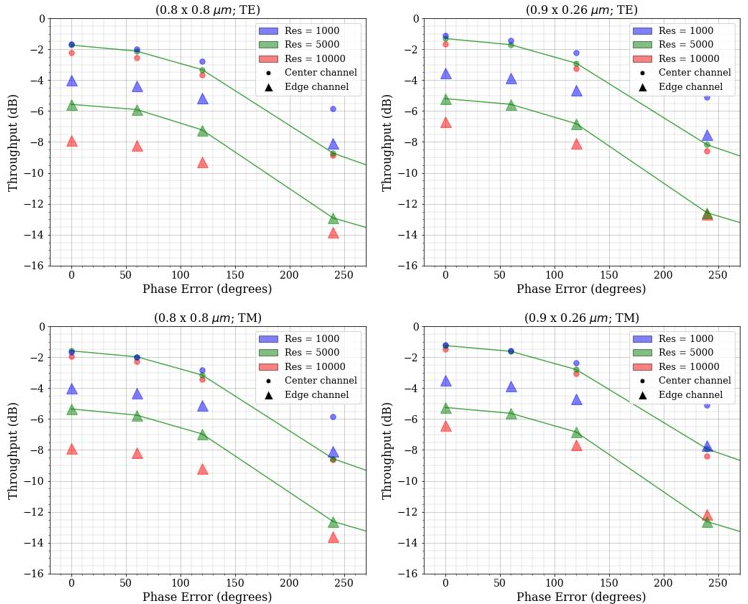}
\caption{\footnotesize Throughput of AWGs with square (Ligentec-like) and rectangular (LioniX-like) waveguides as a function of phase errors, for various resolving powers for both TE and TM polarizations.}\label{fig:Thru_vs_PE}
\end{figure}

Figure~\ref{fig:Thru_vs_PE} shows the
throughput of the AWG devices as a function of the phase error and resolving power, for TE and TM modes, and for both the square and rectangular waveguides. In each plot, we show the throughput for the central (circle) and edge (triangle) channel of the spectral order around 1550 nm. From the plot it can be seen that the throughput of the central channel does not significantly vary with the resolving power. On the other hand, the throughput of the edge channels is substantially lower at higher resolutions. This is likely due to the sidelobes that arise given the asymmetric illumination received at the output edge channels and become more significant at higher resolutions (discussed later in this section). 

With no phase error present the devices could achieve throughputs of $-$1 to $-$2 dB (80$-$63\%). The throughput reduces monotonically as a function of increasing phase error, which as outlined above, is because the light is spatially smeared out across the output of the second FPR and thus less light is coupled into a given output waveguide. It can be seen that with phase errors of up to 120$^{\circ}$ there is minimal additional loss ($<$ 1.5 dB or $<$ 30\%), which is consistent with Figures~\ref{fig:LGT_spectra} and \ref{fig:LiX_spectra} above. To maintain high efficiency, it is critical to minimize phase errors in the array to the greatest extent possible. 

The throughput of the edge channels is lower than the central channel. As explained in Section \ref{subsec:AWG_prop}, this is determined by the far field pattern of the waveguides illuminating the output FPR. It is clear that the square waveguides have a higher loss for the edge channels as compared to the rectangular guides, thus reducing the overall device efficiency. This should be considered in the overall loss budget while designing a photonic spectrograph. 

Finally, the throughput of the AWGs does not appear to vary significantly between the TE and TM modes for both waveguide geometries. 

\begin{figure}[h!]
\centering\includegraphics[width=0.99\textwidth]{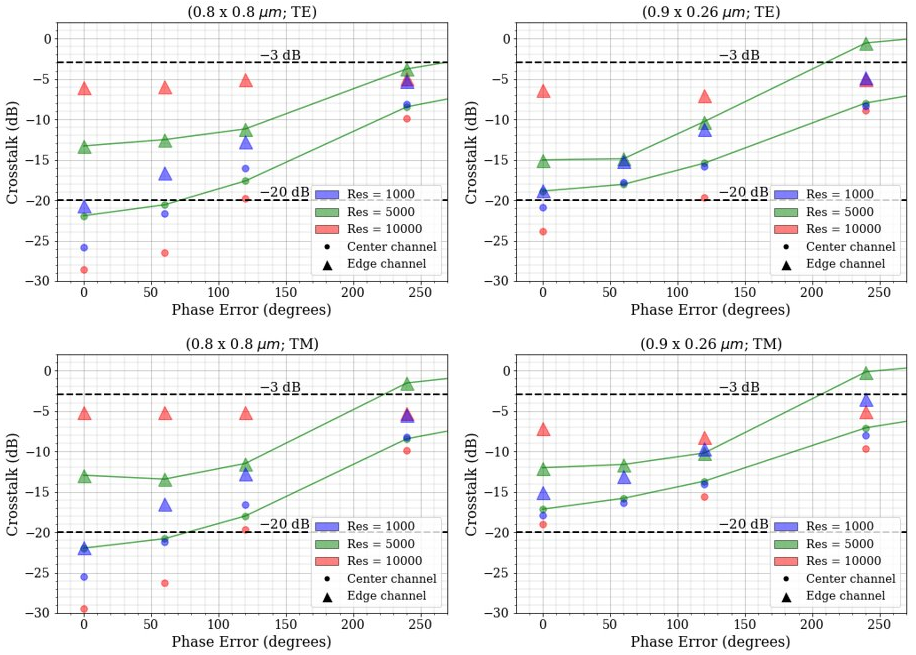}
\caption{\footnotesize Crosstalk of AWGs with square (Ligentec-like) and rectangular (LioniX-like) waveguides as a function of phase errors, for various resolving powers for both TE and TM polarizations.}\label{fig:Xtalk_vs_PE}
\end{figure}

\begin{figure}[h!]
\centering\includegraphics[width=0.99\textwidth]{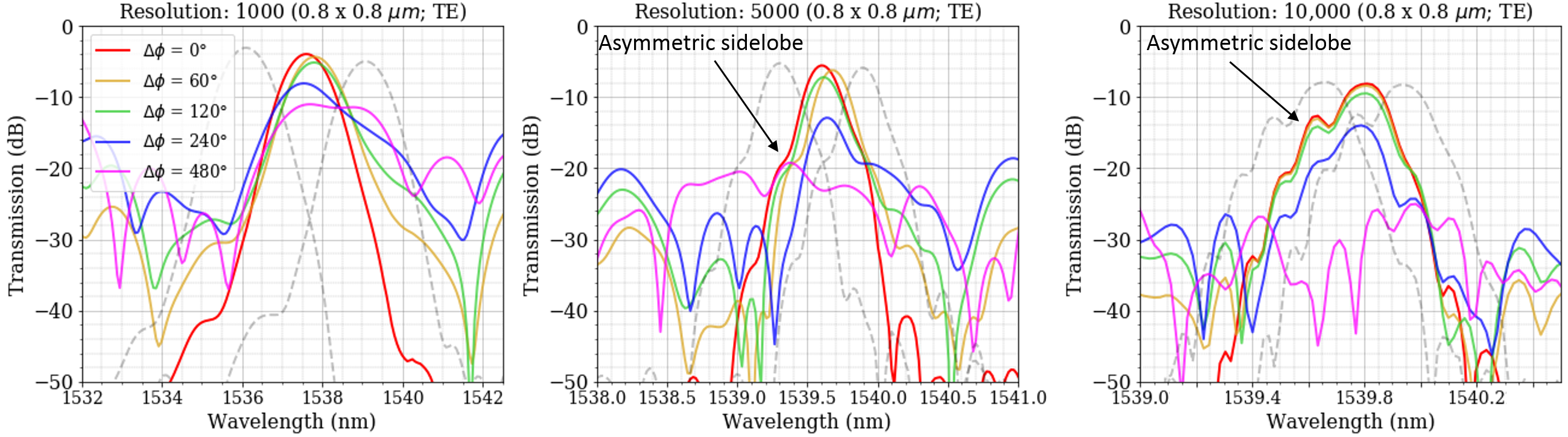}
\caption{\footnotesize Spectral response of a single `edge channel' of the AWG with square waveguides for TE polarization, as a function of resolving power and phase errors. The asymmetric sidelobe arising from the asymmetric illumination of the edge channels is clearly shown. The dotted lines show the neighboring channels at zero phase error.  It can be clearly seen that at higher resolutions, the asymmetric sidelobe dominates the crosstalk.  }\label{fig:sidelobe_edge}
\end{figure}

Figure~\ref{fig:Xtalk_vs_PE} shows the crosstalk of the AWG devices as a function of phase error and resolving power, for TE and TM modes, and for both the square and rectangular waveguides. In each plot, we show the crosstalk for the central and edge channel for a spectral order around 1550 nm. It can be seen that in nearly all cases, the cross-talk increases as a function of increasing phase error. However, for the edge channels, this trend gets flatter at higher resolutions. This is because the edge channels see an additional one-sided sidelobe that arises due to the asymmetric illumination that they receive from the waveguide array. This is shown in Fig. \ref{fig:sidelobe_edge} for the square waveguide in TE polarization. As seen in this figure, this additional sidelobe dominates the crosstalk for the edge channels. The transmission of the peak and the sidelobe degrade in a similar fashion with increasing phase errors, therefore the variation of edge-channel crosstalk appears flatter (as compared to the central channels). 

The dashed $-$3 dB  line (50\% spectral purity) in Fig.~\ref{fig:Xtalk_vs_PE} is a threshold, which indicates one would have a poorly defined spectral channel with $>$ 50\% contamination from the neighboring channels, thus making the device ineffective. The edge channels approach this threshold sooner as compared to the central channels because of the additional asymmetric sidelobes (see Fig. \ref{fig:sidelobe_edge}).

If the sole requirement was to ensure all devices stayed above this threshold, then phase errors should typically be maintained below 120$^\circ$. In this case, <$-5$~dB of crosstalk can be achieved at all resolutions (for both central and edge channels), with both waveguide geometries and in both TE and TM polarizations with phase errors of <$120^\circ$. In general, one will most likely not disperse the output of these waveguide channels any further and hence get a single flux measurement per channel (or per discrete resolution element). In that case, minimizing crosstalk is ideal to minimize the wavelength confusion in reconstruction of the spectrum. If a $-$20 dB crosstalk was achieved, then 99\% of the light in a given channel would be from a specific wavelength range (or a resolution element) alone, thus minimizing wavelength confusion and improving the quality of the final reconstructed spectrum. This level is shown in dotted lines in Fig. \ref{fig:Xtalk_vs_PE}. This goal (for the central channels) could be achieved with <$60^\circ$ of phase error for the square waveguides, but only with $\sim0^\circ$ phase error for the TE polarization when rectangular guides are used.      

\subsection{Spectral drop out}
\begin{figure}[h!]
\centering\includegraphics[width=0.99\textwidth]{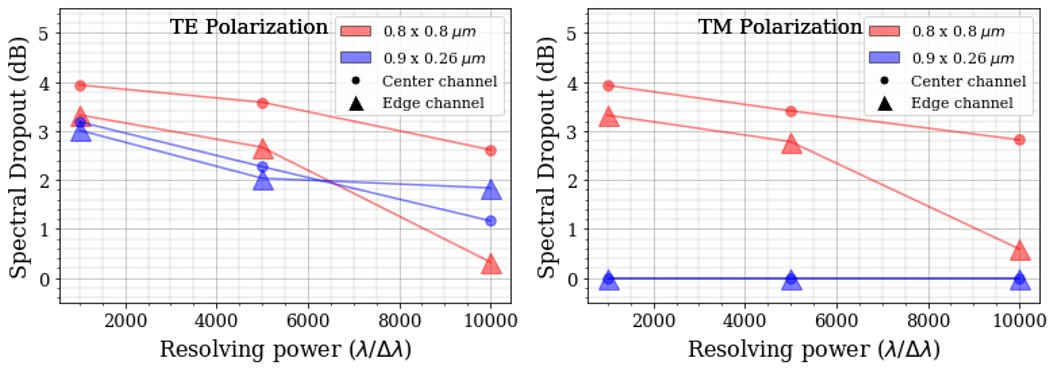}
\caption{\footnotesize Spectral drop-out as a function of resolving power for both the square and rectangular waveguides for TE (Left) and TM (Right) polarizations.}\label{fig:AWG_dropout}
\end{figure}

The spectral drop-out, defined in Section~\ref{subsec:AWG_prop}, as a function of R for both the TE and TM polarizations is shown in Fig.~\ref{fig:AWG_dropout}. For square waveguides, the spectral dropout is similar for TE and TM polarizations, with 3$-$4 dB at R=1,000 dropping monotonically towards 0$-$3 dB at R=10,000. Similarly, for the rectangular waveguides, the dropout decreases monotonically with increasing R in the TE polarization up to R=5,000.
However, in the TM polarization, they show no spectral drop-out. This is primarily due to the broad TM mode profile which leads to a broader spectral profile compared to the TE mode for a given AWG (as shown in Fig. \ref{fig:LiX_spectra}), resulting in a higher crosstalk with the neighboring waveguides, but at the same time, a lower spectral dropout.  

\subsection{Coupling to/from AWGs}
Understanding the waveguide properties, namely the MFD and NA at the facets, is critical to efficiently interfacing the fibers/detectors with the AWGs. 
To efficiently couple to standard NIR SMFs, which have typical MFDs of $\sim$10~$\upmu$m, requires the use of tapers. Two main types of tapers can be used: 1) inverted tapers within the plane of the circuit and 2) 3D tapers (sometimes refered to as `spot size converters'). Inverted tapers can be implemented by reducing the width of the waveguide towards the facet of the chip \cite{zhu2016ultrabroadband}. As the aspect ratio of the waveguide is modified during the taper, the shape of the mode and the polarization properties change in this region of the device. 3D tapers alter the waveguide width and height resulting in a more isotropic expansion of the waveguide mode \cite{mu2020edge}. Another advantage of the 3D tapers is that a larger MFD range can be achieved with better reproducibility (i.e.~the mode can be expanded more) than by using an inverted taper. As such, 3D tapers offer lower losses when coupling to the SMF28 fiber. Besides offering lower losses, they are also less sensitive to mis-alignments. On the other hand, inverted tapers are easier to fabricate (without requiring additional etching steps), and hence are cheaper and widely used. Ligentec offers both types of tapers, but the 3D versions are not standard as part of MPW runs and have an additional cost. For Lionix, both inverted tapers as well as 3D tapers, which can match to the 10~$\upmu$m SMF mode are part of the MPW offering.  

\begin{figure}[h!]
\centering\includegraphics[width=0.99\textwidth]{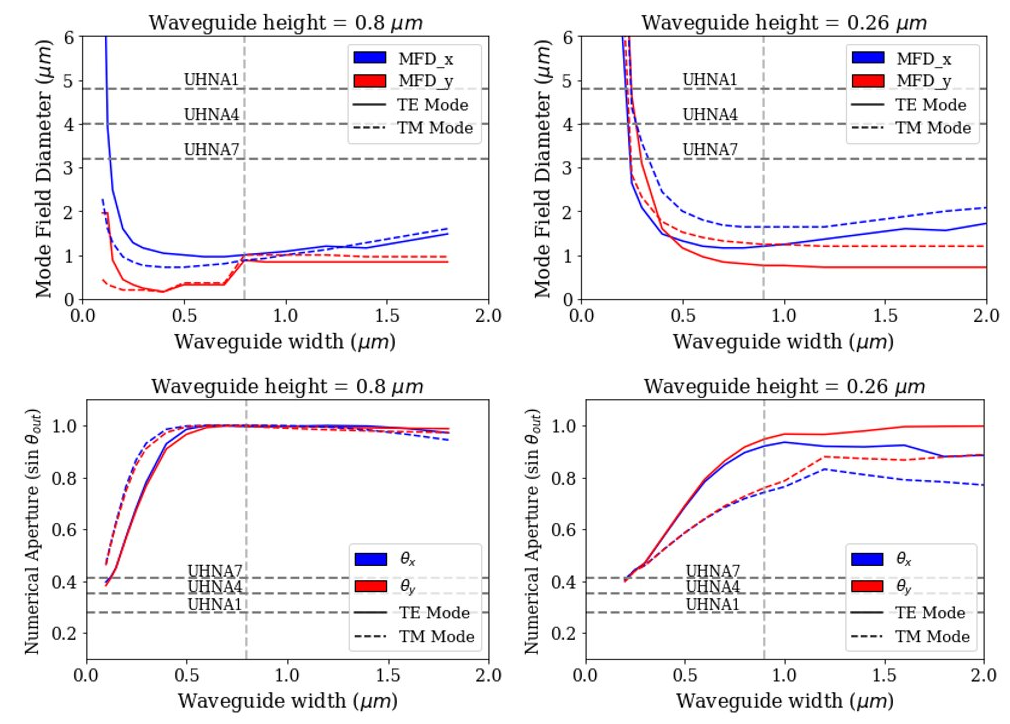}
\caption{\footnotesize MFD and NA versus waveguide width, with a fixed waveguide height, for both square (Ligentec-like) and rectangular (LioniX-like) geometries, for TE and TM polarizations in x and y. }\label{fig:MFD_NA}
\end{figure}

To get a sense of the properties of inverted tapers, Fig.~\ref{fig:MFD_NA} shows the MFD and NA as a function of the waveguide width (for a fixed waveguide height). As outlined above, the inverted taper is formed by shrinking the width of the waveguide (i.e.~the regions to the left of the vertical black dashed lines indicating the nominal waveguide widths utilized throughout this paper). 

Figure~\ref{fig:MFD_NA} shows the range of possible output MFDs/NAs with inverted tapers. As can be seen, at very small widths, the MFD blows up. Given the steepness of the curve at this regime it is extremely difficult to get a precise output MFD. Hence it is challenging  to reliably create an inverted taper that can enlarge the MFD to optimize coupling with SMF28 (although LioniX has done this). Therefore, the ultra-high NA fibers,  which have smaller MFDs can be used as an intermediate step. In this case the UHNA fiber can be spliced to SMF28 and by applying a short taper to the spliced region a low loss conversion can be easily realized. In addition to UHNA fibers, lensed fibers can also be used. The benefit of lensed fibers is that the output MFD is continuously selectable between 2 to 6~$\upmu$m at the time of manufacturing, unlike with the UHNAs which only come in discrete values, allowing for more precise mode matching. Also, the lensed fiber can be fabricated on almost any fiber, which means no splicing taper is needed. One challenge of the lensed fibers is bonding them to the chip as the tip can not be immersed in glue (since the spot is usually formed at some distance from the fiber). 

\begin{figure}[h!]
\centering\includegraphics[width=0.99\textwidth]{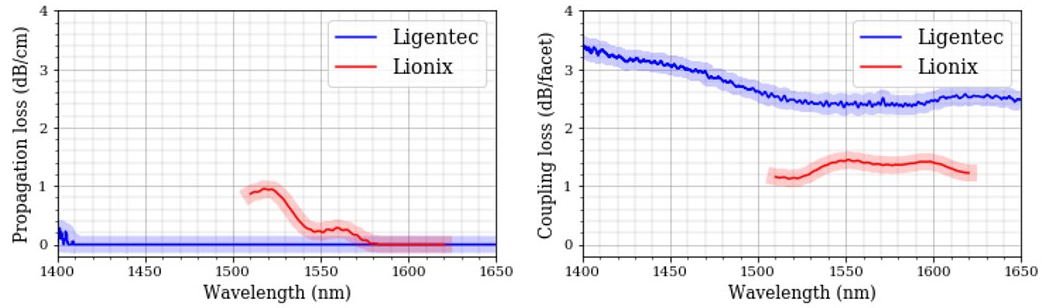}
\caption{\footnotesize Measured propagation and coupling losses to Ligentec and LioniX MPW waveguides for the TE polarization only. The coupling loss is measured with an SMF28 fiber. The fainter shades show the uncertainty in measurements. }\label{fig:losses}
\end{figure}

Figure~\ref{fig:losses} shows the experimentally measured propagation and coupling losses to Ligentec and LioniX (dual stripe) MPW waveguides. The Lionix and Ligentec measurements were done with SMF28 in different labs. The Lionix and Ligentec reference waveguides had inverted tapers for coupling. The propagation losses were difficult to quantify with tight error bars due to their small values (<1 dB/cm) and the limited reference waveguides we had access to. In the future, we will test more devices and refine these values with better strategies for quantifying the propagation loss \cite{hu2018characterization}. The coupling losses for the Ligentec waveguides (with inverted tapers for UHNA4 fiber), are as low as 2.3 dB/facet around 1550 nm, but increase at shorter wavelengths. On the other hand, as a result of the larger MFDs of the LioniX inverted tapers, they can be efficiently coupled to SMF28 directly, with losses <1.5 dB/facet around 1550 nm. A summary of the options for coupling from fibers to waveguides for the two platforms is shown in Table~\ref{tab:coupling}. 

\begin{table}[h!]
\footnotesize
\centering
\begin{tabular}{|c|c|c|}
\hline
\textbf{Coupling type} & \textbf{Square/Ligentec} & \textbf{Rectangular/LioniX}\\
\hline
Inverted taper coupling to SMF28     & 5.4 dB/facet$^{1}$                          &  as low as $0.5$ dB/facet$^{4}$              \\  
Inverted taper coupling to SMF28     & $\sim$2.3 dB/facet$^{2}$ (2 {\rm dB/facet$^{3}$})  &           -          \\  
3D taper coupling  to SMF28          & $<1$ dB/facet$^{3}$                        &  $<0.5$ dB/facet$^{4}$ \\  
\hline
\end{tabular}
\caption{Coupling efficiencies around 1550 nm. 1. Simulated. 2. Measured by the authors on a batch of recent Ligentec waveguides. 3. Value provided by Ligentec. 4. Value provided by LioniX.}\label{tab:coupling}
\end{table} 

Besides coupling to/from fibers, it may be desirable to image the output beam. To do this the optical system must be able to capture all of the light emanating from the waveguide, which is easiest when the NA is $<0.3$, since most inexpensive commercial optics operate below this level. Figure~\ref{fig:MFD_NA} can be used to determine the properties of the inverted taper needed to be able to have a given NA to support the imaging system. If the focal plane (FPR) of the AWG is imaged directly (i.e.~ not sampled by discrete waveguides), then it is difficult to use tapers. This means the output facet is a slab cross-section (of the FPR). 
This will result in the smallest MFD and largest possible output NA, placing extreme demands on the imaging system in the back end. Careful consideration must be given in this case to build a suitable imaging system. .  

\subsection{Footprint}
\begin{figure}[h!]
\centering\includegraphics[width=0.5\textwidth]{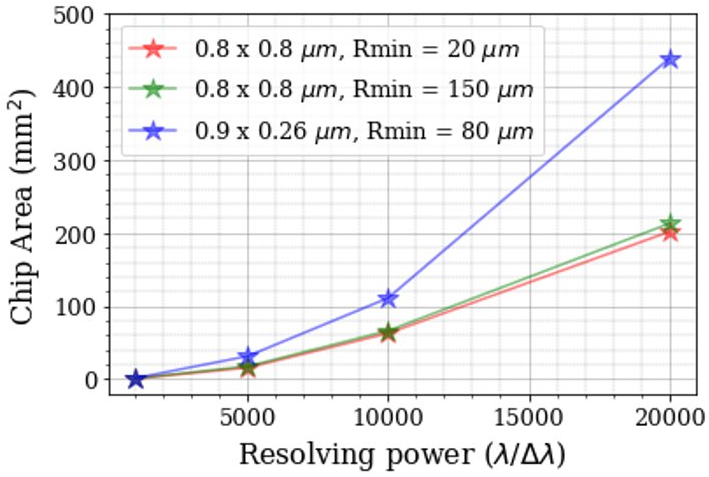}
\caption{\footnotesize Footprint of the AWG chip (i.e.~the area) as a function of resolving power for square and rectangular waveguide geometries considered in this paper. The R$\mathrm{_{min}}$ is the minimum radius of curvature allowed in order to minimize the radiation losses due to bends. For the square geometry (Ligentec-like), R$\mathrm{_{min}}$ = 20~$\mathrm{\upmu m}$ is recommended if only TE mode is of interest and R$\mathrm{_{min}}$ = 150 $\upmu$m is recommended to support both polarizations (to minimize the bend losses). }\label{fig:footprint}
\end{figure}

Figure~\ref{fig:footprint} shows the area of the device as a function of the resolving power and the waveguide geometry. It is clear that the area grows monotonically with increasing R, which is because both the number of waveguides and the incremental path length in the array increase with R, thus requiring a larger footprint. For the square waveguide, it can be seen that using tighter bend radii as low as 20~$\upmu$m brings only marginal gains compared to a device with bend radii $>150~\upmu$m. However, square waveguides (eg: 0.8 $\times$ 0.8 $\upmu$m), which produce smaller modes, offer significant advantages in terms of footprint over larger modes provided by rectangular waveguides (eg: 0.9 $\times$ 0.26 $\upmu$m), particularly at R > 10,000. The area of a reticle offered by Ligentec is 10$\times$5 mm$^2$ versus LioniX which offers 16$\times$16 mm$^2$. Based on these cell sizes, the plot indicates that an R=10,000 device could fit onto a Ligentec cell, while a R$\sim$15,000 device could fit within a LioniX cell. It should be noted that there are more optimal ways to arrange AWG circuits which could be used to push these limits, but this gives a sense of what can be accommodated within a reticle. Another thing to note is that larger devices are more susceptible to process variation and accumulating phase errors and this could be the dominant error source limiting the performance of the device. This should be carefully considered with the foundry and tested experimentally.  

\subsection{Phase errors and fabrication tolerances} \label{subsec:phase_err_fabrication}
In order to put the simulation results described below into perspective, it is important to understand how the fluctuations in the width and height of a waveguide lead to phase errors in an AWG. In our treatment of phase errors, we have assigned a randomly selected phase error for each waveguide in the array ($\Delta\phi_{\rm wg}$), and thereby a randomly selected $\Delta n_{\rm eff, wg}$. In other words, each waveguide in the array is assumed to have a constant (but randomly selected) error in the effective index ($\Delta n_{\rm eff, wg}$ along the entire length of the waveguide, $L_{\rm wg}$). In reality the error in the effective index of a waveguide will vary along the length of the waveguide, with some parts of the waveguide having a larger fluctuation than the others, thus giving rise to an equivalent net error in the $n_{\rm eff}$ of that waveguide. Hence, the current treatment is a reasonable approximation to study the effect of total phase errors. The following equations show how the phase error in angle ($\Delta\phi_{\rm wg}$) relates to the error in $n_{\rm eff}$, which in turn depends on the fluctuations in waveguide width or height.

\begin{equation}
   \phi_{\rm wg} = \frac{2\pi L_{\rm wg}}{\lambda_{0}/n_{\rm eff}}\\
 \end{equation}

\begin{equation} \label{eqn:delta_phi}
\therefore \Delta\phi_{\rm wg} = \left(\frac{2\pi L_{\rm wg}}{\lambda_{0}}\right) \Delta n_{\rm eff, wg}
\end{equation}

It was previously shown experimentally in silica platforms that the phase errors grow linearly with the length of the waveguides \cite{stoll2020performance}. If we convert the $\Delta n_{\rm eff, wg}$ to $\Delta W_{\rm wg}$ (fluctuation in the width of the waveguide) assuming a fixed height or, $\Delta H_{\rm wg}$ (fluctuation in the height) assuming a fixed width, we get an estimate of the allowable fabrication tolerances for a given AWG. Note that this conversion depends on the fractional length of the waveguide affected by such a fluctuation (dictated by the $L_{\rm wg}$ term in equation \ref{eqn:delta_phi}). Assuming  $\Delta\phi_{\rm wg}$ = 180$^{\circ}$ in equation \ref{eqn:delta_phi}, we first obtain the $\Delta n_{\rm eff, wg}$. Then using the effective index calculations of the waveguide geometries as a function of width and height (for the TE mode), we calculate the relevant $\Delta W_{\rm wg}$ or $\Delta H_{\rm wg}$ that would give rise to the corresponding $\Delta n_{\rm eff, wg}$.

\begin{figure}[h!]
\centering\includegraphics[width=0.99\textwidth]{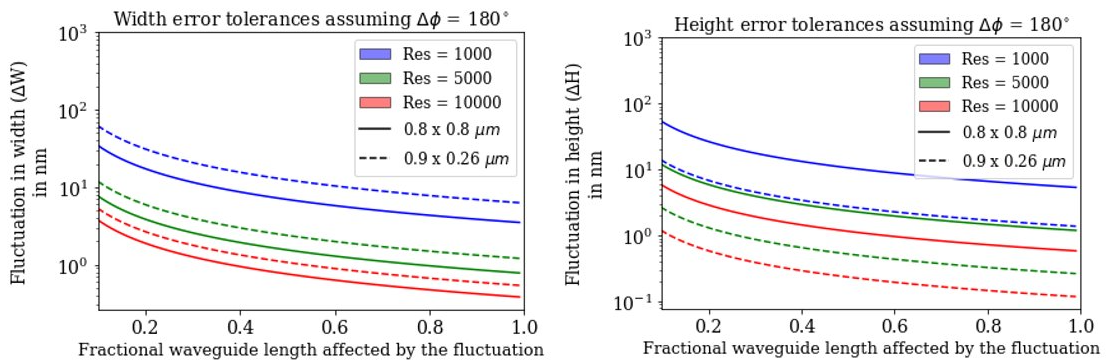}
\caption{\footnotesize Magnitude of the fluctuations in the width (Left) and height (Right) of the waveguide as a function of the length of the fluctuation that would induce a 180$^{\circ}$ of error. Square waveguides are shown in solid lines, while rectangular waveguides are highlighted with dashed lines.  }\label{fig:phase_error_vs_irrergularities}
\end{figure}

Figure~\ref{fig:phase_error_vs_irrergularities} shows what fluctuation in the waveguide width (left panel) or height (right panel) could be tolerated for AWGs with various resolving powers designed with both waveguide geometries, as a function of the fractional length of the fluctuation (= section of the waveguide affected by the fluctuation / total length of the waveguide) to achieve an overall maximum phase error of 180$^{\circ}$. A fractional length fluctuation of 1 means that the waveguide width or height error uniformly affects the entire length, whereas smaller values mean that the error is localized to only a section of the waveguide. This error only refers to the waveguides in the array.

It is clear from Fig. \ref{fig:phase_error_vs_irrergularities} that the shorter the fractional length of the width or height fluctuation, the larger the tolerance to keep the phase error under a given threshold. The figure shows that lower resolving power AWGs can handle larger width or height fluctuations by up to an order of magnitude. This is a result of the waveguides in the array being longer for higher resolution devices, which makes them more sensitive to fluctuations in dimensions. 

In addition, the rectangular waveguides (0.9 $\times$ 0.26 $\upmu$m) are more tolerant to width variations compared to the square waveguides (0.8 $\times$ 0.8 $\upmu$m), while being less tolerant to height variations, as one would expect given their aspect ratio. As an example, with a  phase error threshold of 180$^{\circ}$, width errors of 10 nm and 1 nm could be tolerated $-$ assuming the entire length of the waveguides (in the array) was affected $-$ for devices with R=1000 and 10000, respectively. This corresponds to a fractional width error ($\Delta W/W$) of 1.3\% and 0.13\% for the square waveguides,  respectively. However, if the defects were localized to say 1/10$^{\rm th}$ of the waveguide length, then the tolerated waveguide width variation would be 40-50~nm and 4-5~nm respectively for R=1000 and 10000. The allowable fractional width errors in this case have increased to 5.6\% and 0.56\%, respectively. Keeping the width fluctuation below 4-5 nm within a reticle ($\sim$ 100 mm$^2$) is achievable with the Ligentec\footnote{private communication,  MPW manual available from Ligentec upon request.}, process for its square waveguides. Similarly, in terms of height, there is a slow variation $\sim $3~nm in the waveguide thickness across a 400 mm$^2$ span of a typical Ligentec wafer. For an AWG with R=10000, which has a typical footprint of 30 mm$^2$ ($\sim6\times6$~mm), a thickness fluctuation of less than 1~nm can be achieved, thus producing a phase error of less than 120$^{\circ}$ (from Eqn.~\ref{eqn:delta_phi} and Fig.~\ref{fig:phase_error_vs_irrergularities}).

Note that better fabrication processes are available in the dedicated runs with the commercial foundries with better optimized tolerances. These processes are preferable for building higher resolution AWGs (R $>$ 10000) which also have larger footprints ($>$ 100 mm$^2$, from Fig. \ref{fig:footprint}).







\section{Discussion}\label{sec:discussion}

To summarize, we presented a scan of AWG spectral characteristics as a function of resolution, phase errors, and polarization for two distinct waveguide geometries used by commercial SiN platforms. While the presented designs do not make use of any specialized treatment to optimize a given spectral characteristic of interest (say, throughput or crosstalk), they provide the community a realistic insight, particularly relevant to astronomical spectroscopy, prior to making fabrication decisions. 

\subsection{Overall Throughput}
As astronomy is a photon-starved field, high throughput is the key requirement. The losses in an AWG spectrograph consist of the following terms: fiber-to-chip coupling (can be tackled by mode matching to a large extent), Fresnel reflection losses (can be tackled by index matching and anti-reflective coating), propagation and bending losses, insertion loss, and chip-to-free-space loss (in the case of detector imaging of the spectrum; need to match NA of the imaging optics). 

The results presented above can be combined with vendor/lab data to get a sense of overall throughput. Given the complexity of different factors giving rise to the phase errors, we can only make a crude estimate of the phase errors  one might expect. As discussed at the end of Sec \ref{subsec:phase_err_fabrication}, for an AWG with R=5000 (typical footprint of 30 mm$^2$), the expected height fluctuation would be less than 1~nm for the  Ligentec MPW process, thus producing a phase error of less than 120$^{\circ}$ (from Eqn.~\ref{eqn:delta_phi} and Fig.~\ref{fig:phase_error_vs_irrergularities}). Conservatively, we assume a 120$^{\circ}$ maximum phase error in the waveguide array to determine the overall throughput for a realistic device.  

Table~\ref{tab:AWG_loss_table} shows the losses of each element of an AWG (assuming R$\sim$5000) along with the total loss, for TE polarization only. The results for both the square and rectangular designs are included along with values for optimized devices with future developments (or existing techniques available in dedicated fabrication runs of the commercial foundries). It can be seen that with the current state of technology, that AWGs based on square waveguides shown here would have losses of 5.9 dB (a throughput of 26$\%$), while rectangular devices would have losses of 4.7 dB (a throughput of 34$\%$) for the TE polarization. These values are a little low, but note that these are estimates for an MPW run without any special optimizations. Significant gains can be made by currently available optimizations in the dedicated runs. For instance, Ligentec offers specialized tapers (e.g.~3D tapers) at the fiber-waveguide interface to match the mode of a SMF28 fiber (MFD: 10~$\upmu$m). This immediately reduces the coupling losses to $<$0.5 dB/facet. By simply improving the phase errors from 120$^{\circ}$ to 60$^{\circ}$, which is achievable with the current commercial platforms, the insertion losses are suppressed by nearly 1.25 dB  (see Fig. \ref{fig:Thru_vs_PE}). The insertion losses can be further improved by optimizing the FPR-waveguide tapers. An improvement in taper efficiency from the 92\% for the linear tapers to 98\% for optimized tapers can further reduce the insertion losses by 1 dB (since there are four such interfaces). We revisit the areas where throughput can be improved in the future in section \ref{sec:future}.

As shown in Table \ref{tab:AWG_loss_table}, a throughput of $\sim 60\%$ can be obtained for the AWG chip. For an estimate of the total throughput on a telescope, we take the case of SCExAO (Subaru Coronagraphic Extreme Adaptive Optics, installed on an 8-m telescope) feeding an SMF which then illuminates the photonic spectrograph \cite{jovanovic2015}.  Assuming an injection throughput of $\sim 45\%$ (from sky to the SCExAO focus)  as described in \cite{jovanovic2017demonstration} and a capture throughput from the focal spot to the SMF of $\sim$ 50\% (corresponding to a Strehl ratio of 65\% in H-band), the total sky-to-detector throughput of the instrument will be $\sim 0.45\times0.50\times0.6$ = 13.5\%. Thus, with these improvements in the AWG, the optimized throughput of a photonic spectrograph is poised to be comparable to the current bulk optics near-IR spectrographs (typically $\sim$15\% sky-to-detector throughput which depends on the specific instrument and configuration \cite{GNIRS,MOIRCS}).  

We remind the reader that these values represent the throughput estimate at the peak of the FSR and that there will be additional losses at the edges of the orders. Further efforts are required to engineer the waveguide structures at the FPR-waveguide interface to suppress this extra loss. In addition, due to discretization of the output field into waveguides, there is also spectral drop-out not included in the table. However, there is no spectral drop-out if the output waveguides are removed and the spectrum is imaged directly onto a detector from the output FPR (eg. see \cite{cvetojevic2012-FSS}).     

\begin{table}[ht]
\footnotesize
\begin{center} 
\begin{tabular}{|C{1.5cm}|C{1.7cm}|C{1.7cm}|C{1.7cm}|C{1.7cm}|C{2.2cm}|} 
\hline
\rule[-1ex]{0pt}{3.5ex}  \textbf{Loss term} & \textbf{Square (dB)} & \textbf{Optimized (dB)} & \textbf{Rectangular (dB)} & \textbf{Optimized (dB)} & \textbf{Improvement approach}\\
\hline
\rule[-1ex]{0pt}{3.5ex}  Insertion loss &  3.2 (48\%) & 1.2 (76\%) & 3.0 (50\%) & 1.0 (79\%)  & Phase-error minimization; Array-FPR taper optimization\\
\hline
\rule[-1ex]{0pt}{3.5ex}  Coupling loss & 2.5 (56\%) & 1.0 (79\%) & 1.5 (71\%) & 1.0 (79\%) & 3D tapers \\
\hline
\rule[-1ex]{0pt}{3.5ex}  Propagation loss (over 1 cm) & 0.2 (95\%) & 0.2 (95\%) & 0.2 (95\%) & 0.2 (95\%) & Already optimized  \\
\hline
\rule[-1ex]{0pt}{3.5ex}  \textbf{Total loss} & 5.9 (26\%) & 2.4 (58\%) & 4.7 (34\%) & 2.2 (60\%) &   \\
\hline
\end{tabular}
\caption{\label{tab:AWG_loss_table} \footnotesize Components of losses in a R $=$ 5000 AWG  shown in dB-scale. The numbers in parentheses show the corresponding throughput for each term in percentages. The numbers under square and rectangular show the losses assuming a phase error of 120$^{\circ}$ and no specific optimizations for slab-waveguide tapers and with the coupling loss measured using simple 2D tapers. The optimized columns show the improved losses after improving the phase errors to  60$^{\circ}$ and incorporating optimized tapers for fiber-waveguide (3D spot-size converters) and waveguide-slab (i.e.~waveguide-FPR) interfaces.}
\end{center}
\end{table}

\subsection{Fabrication tolerances}
In terms of tolerances, it is clear that tighter process control is required for high-resolution devices. Due to the longer lengths of the waveguides in the array required for the high-resolution AWGs, they are  more susceptible to fluctuations in the height or width dimensions. Therefore, the larger the footprint of the device, the greater the susceptibility and hence tighter the required tolerances. 

From Figs. \ref{fig:LGT_spectra}-\ref{fig:Xtalk_vs_PE}, it is clear that we should not exceed the $\Delta\phi$ beyond 180$^{\circ}$ in any case, but even tighter tolerances can help with improved throughput and reduced cross-talk. Despite that fact we have demonstrated that AWGs with a range of resolutions can be designed, but manufacturing limitations, which induce phase errors and degrade performance, need to be carefully understood to determine what can be realized. Although some properties such as uniformity of waveguide thickness and width versus location on a wafer can be measured and fed into simulations, ultimately devices need to be fabricated to fully assess the actual phase errors that are present. This is where a close working relationship with the foundry, and several iterations of device design, fabrication and characterization will be needed. Indeed for best results, the foundry should be involved from the beginning of the design stage.

\subsection{Polarization sensitivity}
Since most astronomical sources are unpolarized, a polarization insensitive spectrograph is preferred to eliminate the need for polarization splitting. 
We explored the throughput and crosstalk characteristics of AWGs with both the square and rectangular geometries. The insertion losses are not significantly different between the two polarizations in either geometries. The crosstalk is higher in TM mode for the rectangular waveguides, but still within acceptable limits for astronomical spectrographs. 

However, the most important issue is the birefringence. The shift in the spectrum due to birefringence is given by : ($n_{\rm eff, TE}~ - ~n_{\rm eff, TM}$)/$n_{\rm eff, mean}$ $\times$ $\lambda$. Due to a nearly 4\% variation in the effective index between TE and TM mode in the rectangular waveguides (from Fig. \ref{fig:mode_profiles}), they can only be used for a single polarization. For square waveguides, the difference is only 0.015\%, thus giving a wavelength shift of 0.23 nm, which is acceptable for R = 1000 ($\Delta\lambda$ = 1.55 nm) and 5000 ($\Delta\lambda$ = 0.31 nm), but for R=10,000 ($\Delta\lambda$ = 0.155 nm) the offset exceeds the bandwidth of a single resolution element and hence will lead to a degradation in the spectral resolution. Therefore, it is more suitable to split the polarization and acquire separate spectra for TE and TM polarizations for high-resolution AWGs (R $>$ 10,000). Alternatively, ultra-low birefringence waveguides could be constructed, which would however require a tighter fabrication control to maintain the low birefringence. Nonetheless, for low-resolution spectrographs, polarization insensitive AWGs can be constructed using square waveguides. 


\section{Future outlook}\label{sec:future}
From Table~\ref{tab:AWG_loss_table} it is clear that the losses of current generation SiN devices are higher than desirable for astronomy. The table also includes columns that show the expected losses/throughput if critical elements of the circuit were optimized (e.g.~coupling). Broadly speaking, these optimizations could include optimized coupling tapers for fiber-waveguide interfaces (such as 3D spot-size converters), improved taper designs to minimize losses at the four waveguide-FPR (i.e.~slab-to-waveguide) interfaces on the chip, and minimizing phase errors by exploring tolerant designs as well as choosing optimized fabrication processes. By exploiting these approaches, AWGs based on square waveguides could have losses of 2.4 dB (a throughput of 58$\%$), while rectangular devices could have losses of 2.2 dB (a throughput of 60$\%$) for the TE polarization. 

A significant contribution to the loss is associated with the insertion loss element, which could be addressed in the future (apart from improving phase errors) with innovative design methodologies such as inverse design to acquire high efficiency over a broad band \cite{sideris2019ultrafast}. Inverse design is based on setting up a metric, for example to maximize throughput, and then allowing the algorithm to explore the parameter space of waveguide widths, heights, and lengths of smaller elements of the taper to optimize the design for the metric. This approach can result in non-intuitive architectures, which can provide superior performance (as demonstrated in \cite{sideris2019ultrafast}). It may be possible to use these design approaches to engineer waveguide cross sections at the waveguide-FPR interfaces to flatten the envelope of AWG peaks in a spectral order (see Fig. \ref{fig:AWG_spec_param}A), thus reducing losses towards the edges of a spectral order. 

In addition, optimizing spectral drop-out is also critical. This term could be optimized through shaping of the input field fed to the first FPR to be, for instance, similar to a box profile, which will create flatter peak profiles at the output waveguides (since the output field profile is an image of the input). Methods such as using parabolic tapers \cite{okamoto1996flat} or 1$\times$2 MMIs \cite{amersfoort1996passband} can be used at the input waveguides to achieve this. Another alternative would be to modulate a sinc profile on the illumination of the waveguides in the array to create a flat (rectangular) wavelength response for each spectral channel (instead of a Gaussian) \cite{okamoto1995arrayed}, since the field profile at the output waveguide is a Fourier transform of the illuminating field from the array.  Both of these methods will effectively reduce the spectral drop-out without significantly increasing the crosstalk (see Sec 9.5.1 in \cite{okamoto2010fundamentals} for more details). 



Currently the fabrication of astrophotonic AWG spectrographs with Ligentec and Lionix MPW processes is underway. These tests will further inform on the manufacturing limitations and the specific improvements that need to be undertaken to enhance the spectral properties of these AWGs.   







\section{Conclusion}
In this paper, we explored the key structural and spectral properties of AWG spectrographs  using commercial MPW geometries from an astronomical perspective. We discuss a range of concerns critical to astronomy including the throughput, injection and imaging of light from the chip, and polarization sensitivity. We also discussed realistic limitations of fabrication by studying the effect of phase errors and exploring how high a resolution spectrograph can be constructed using commercial MPW processes. We find that R $\sim$ 10,000 are easily achievable in the commercial MPW runs and reiterate that more dedicated runs will be able to provide greater fabrication stability and tolerances, and thereby the ability to construct even higher resolution AWGs without accumulating phase errors $>$ 180$^{\circ}$. We provide a breakdown of loss terms in an AWG spectrograph and specify avenues $-$ coupling loss, FPR-waveguide tapers and advanced designs $-$ where greatest improvements can be made. We also estimate the expected AWG throughputs through currently available optimized strategies at these foundries (in dedicated fabrication runs). The expected throughput from these strategies ($\sim$ 60\%) would position these devices to be competitive with the classical astronomical spectrographs. Finally, we provide a future outlook for the astrophotonics community on the thrust areas to make astrophotonic spectrographs an attractive alternative to the classical astronomical spectrographs.

\begin{backmatter}
\bmsection{Acknowledgments}
Pradip Gatkine was supported by the David and Ellen Lee Postdoctoral Fellowship at the California Institute of Technology. Simon Ellis acknowledges the help and expertise of LioniX International in the fabrication of one of the chips tested in this work as part of their MPW service. Nemanja Jovanovic acknowledges the help and expertise of BRIGHT Photonics and Ligentec in the design and fabrication of one of the chips tested in this work as part of their MPW service. This work was supported by the Wilf Family Discovery Fund in Space and Planetary Science, funded by the Wilf Family Foundation. This research was carried out at  the California Institute of Technology and the Jet Propulsion Laboratory under a contract with the National Aeronautics and Space Administration (NASA) and funded through the President's and Director's  Research $\&$ Development Fund Program.

\bmsection{Disclosures}
The authors declare no conflicts of interest.

\end{backmatter}


\bibliography{main}

\end{document}